\documentclass[twocolumn,showpacs,preprintnumbers,amsmath,amssymb,superscriptaddress,floatfix]{revtex4}

\usepackage{graphicx}
\usepackage{latexsym}
\usepackage{epsfig}
\usepackage{textcomp}

\begin{document}
\title{Measurement of the hyperfine splitting of the 6S$_{1/2}$ level in rubidium.}
\author{A. P\'{e}rez Galv\'{a}n, Y. Zhao, L. A. Orozco}
\affiliation{Joint Quantum Institute, Department of Physics, University of Maryland and
National Institute of Standards and Technology, College Park, MD 20742-4100,
USA.}
\date{\today}

\begin{abstract}
We present a measurement of the hyperfine splitting of the
6S$_{1/2}$ excited level of rubidium using two photon absorption
spectroscopy in a glass cell. The values we obtain for the magnetic
dipole constant $A$ are 239.18(03) MHz and 807.66(08) MHz for
$^{85}$Rb and $^{87}$Rb, respectively. The combination of the magnetic moments of the two isotopes and our measurements show a hyperfine anomaly in this atomic excited state. The
observed hyperfine anomaly difference has a value of
$_{87}\delta_{85}=-0.0036(2)$ due to the finite distribution of
nuclear magnetization, the Bohr-Weisskopf effect.
\end{abstract}

\maketitle
\section{Introduction}

Precise measurements of hyperfine splittings of the ground and
excited states are necessary to explore the complete dynamics of the
electron cloud-nucleus interaction in the atom.
New experimental approaches such as femtosecond frequency combs and
small linewidth lasers together with laser cooling and
trapping reach now increased accuracy for high precision studies of
hyperfine structure in excited levels
\cite{barwood91,gerginov03,marian04,marian05,chui05}. This new wave
of experiments has renewed the interest of theorists in predicting accurate electron-nucleus interactions. These calculations of hyperfine splittings in excited states, where electron correlations are less complicated, are more sensitive to nuclear structure details \cite{dzuba05}.

Measurements of hyperfine splittings are also of interest to the
atomic parity non-conservation (PNC) community. Experiments of
atomic PNC rely heavily on calculations of operator expectation
values to extract from the experimental data information on the weak
interaction. The accuracy of the calculations is gauged against
expectations values of atomic properties such as energy levels,
ionization energy, electric dipole operators of the electronic
levels, fine and hyperfine splittings. The hyperfine splitting
measurements, in particular, represent ideal benchmarks for the
\textit{ab initio} calculations of the electronic wave function at
distances close to the nucleus \cite{gomez04,grossman00a}.
Currently, the PNC \textit{ab initio} calculations of theoretical groups
using many body perturbation theory (MBPT) have reached a precision better than 1\% \cite{safronova99,ginges04,johnsonbook2}.

We present in this paper the details of the measurement of the
hyperfine splitting of the 6S$_{1/2}$ excited level in $^{85}$Rb and
$^{87}$Rb \cite{perez07}. We perform the experiment in a glass
cell with rubidium vapor with natural isotopic abundances under a controlled environment. We observe in our experimental data deviations from the assumed point interaction between the valence
electron and the nucleus \textit{i.e.} a hyperfine anomaly, and find that a change in the distribution of the nuclear magnetization between isotopes explains the observation.

The organization of the paper is as follows: section II gives the
theoretical background, section III contains the methodology and
experimental setup as well as the studies of possible systematic
errors and results. Section IV shows the comparison with theory and
section V contains the conclusions.

\section{Theoretical background}

Pauli suggested in 1924 that the ``hyperfine'' splittings observed
in very precise spectroscopic studies of fine structure in atoms
were due to the interaction between the atomic electrons and the
magnetic moment of the nucleus \cite{pauli}. Three years later Back
and Goudsmit succeeded in analyzing the very small splittings of
bismuth using the assumption of Pauli and the coupling of angular momenta
\cite{back27,back28}. Theoretical predictions of the
splittings were moderately successful in explaining the
size of the experimental findings, mainly due to the complexity of
the relativistic many-electron system. However, in the last twenty
years there has been an exceptional output of very accurate
theoretical results coming from MBPT culminating in the extraction
of weak interaction couplings from the atomic PNC measurement in Cs
\cite{wood97,wood99}. There are currently proposals to perform a complementary
PNC measurement in francium where calculations of atomic properties are reaching the 
precision of those in cesium \cite{gomez07}.

\subsection{Hyperfine splitting}

Although a complete treatment requires a full relativistic theory,
estimations of the interplay between the nuclear moments and the
electromagnetic fields created by the electron following classical
electrodynamics agree with the experimental results and
provide physical insight of the phenomenon (see for example
Refs.\cite{kopfermann, corney}). We follow this approach in the
discussion below. The interested reader should consult Ref.
\cite{armstrong} for a relativistic derivation.

Two types of nucleus-electron interactions suffice to account for
the hyperfine splitting in most atoms. The largest of the
contributions comes from the nuclear magnetic dipole coupling to the
magnetic field created by the electrons at the nucleus. The second
one arises from the interaction between the nuclear electric
quadrupole and the gradient of the electric field generated by the electrons at the nucleus. The latter
vanishes for spherically symmetric charge distributions which
correspond to electronic angular momentum $J$ equal  to $0$ or
$1/2$. The hyperfine energy shift $E_{HF}$ for these levels is
\cite{corney}:
\begin{equation}
E_{HF}=\frac{A}{2}(F(F+1)-I(I+1)-J(J+1)),
\end{equation}
where $F$ is the total angular momentum, $I$ is the nuclear spin and
$A$ is the magnetic dipole interaction constant. The derivation of
$A$ for a hydrogen-like atom by Fermi and Segr\`{e} assumes a point nuclear
magnetic dipole \cite{kopfermann}
\begin{equation}
A_{point}=\frac{16\pi}{3} \frac{\mu_{0}}{4\pi h}g_{I}\mu_{N}
\mu_{B}|\psi(0)|^{2}, \label{magnetic dipole constant equation}
\end{equation}
where $\psi(0)$ is the electronic wave function evaluated at the
nucleus, $\mu_{B}$ is the Bohr magneton, $\mu_{N}$ is the nuclear
magneton and $g_{I}$ is the nuclear gyromagnetic factor.

The nuclear magnetic dipole acquires an extra potential energy under
an external DC magnetic field. For small values of the field
($g_{F}\mu_{B}B/E_{HF}\ll1)$ $F$ is a good quantum number and the
energy of the system is given by
\begin{equation}
E_{HF}(B)=E_{HF}(0)+g_{F}\mu_{B}m_{F}B,
\end{equation}
where $g_{F}$ is the total g-factor, $m_{F}$ is the magnetic quantum
number, $B$ is the magnetic field  and E$_{HF}(0)$ is the value of
the energy at zero magnetic field. In this regime of small
splittings compared to E$_{HF}(0)$, $g_{F}$ is given by:
\begin{eqnarray*}
\lefteqn{g_{F}=g_{J}\frac{F(F+1)+J(J+1)-I(I+1)}{2F(F+1)}~-}&\\
  & & g_{I}\frac{F(F+1)+I(I+1)-J(J+1)}{2F(F+1)},\nonumber\\
\end{eqnarray*}
where $g_{J}$ is the electronic $g$-factor.

\subsection{\textit{Ab initio} calculations}

Hyperfine interactions represent a formidable many body problem. A
thorough study must approach the problem from a relativistic
standpoint which further complicates the interactions in a
multi-electron atom. In recent years relativistic MBPT has shown
itself to be a powerful and systematic way of extracting, from the
high quality wave functions that it generates, precise atomic
properties such as hyperfine splittings
\cite{johnsonbook2,safronova04}.

The full method is outlined in Ref.\cite{safronova98} and references
therein. Briefly, the method, applied to alkali atoms, consists of
evaluating a no-pair relativistic Hamiltonian with Coulomb
interactions with a frozen core Dirac-Hartree-Fock wave function of
a one valence electron atom. The Hamiltonian includes projection
operators to positive energy states of the Dirac Hamiltonian. Their
presence gives normalizable, bound state solutions. The wave
function contains single and double excitations to all orders; these
correspond to wave functions useful for calculating energy levels
and transition matrix elements. In order to calculate accurate hyperfine
constants a set of triple excitations has to be added. The
evaluation of the wave function yields coupled equations that are
solved iteratively for the excitation coefficients which are then
used to obtain atomic properties.

The calculations of the hyperfine constants are corrected for the
finite size of the nuclear magnetic moment up to zeroth order only
due to their small size in the lighter alkalies (Na, K, Rb). In
cesium and francium the correction becomes more important and is
included to all orders. The calculation ignores isotopic changes of
the magnetization distribution and it is modeled as a uniformly
magnetized sphere for all the atoms. The magnetization radius used
is equal to the charge radius and the neutron skin contribution is
ignored.

\subsection{Hyperfine anomalies}

The atomic electron sees the nucleus, most of the time, as a
structureless entity with a single relevant parameter, its charge
$Z$. We should expect the electronic wave functions of different
isotopes, to a very good approximation, to be the same. It follows
then, using Eq. \ref{magnetic dipole constant equation} that
\begin{equation}
\frac{A^{87}_{point}}{A^{85}_{point}}=\frac{g^{87}_{I}}{g^{85}_{I}},\label{pointinteractionequality}
\end{equation}
where the superindex denotes the atomic number of the isotope.

However, high precision experiments show differences or anomalies
from this description. It is necessary to consider the nucleus as an
extended, structured object with specific finite magnetization and
electric charge distributions for each isotope. We can express the
anomaly by writing the magnetic dipole constant of an extended
nucleus $A_{ext}$ as a small correction to $A_{point}$
\cite{kopfermann}

\begin{eqnarray}
A_{ext} & = & A_{point}f_{R}(1+\epsilon_{BCRS})(1+\epsilon_{BW}),\nonumber \\
\end{eqnarray}
where $f_{R}$ represents the relativistic correction. The last two
terms in parenthesis modify the hyperfine interaction to account for
an extended nucleus. The Breit-Crawford-Rosenthal-Schawlow (BCRS)
correction \cite{rosenthal32,crawford49,rosenberg72}, the largest of the two,
modifies the electronic wave function inside the nucleus as a
function of the specific details of the nuclear charge distribution.
The second one, the Bohr-Weisskopf (BW) correction \cite{bohr50},
describes the influence on the hyperfine interaction of finite space
distribution of the nuclear magnetization.

Direct extraction of hyperfine anomalies from the experimental data
requires theoretical knowledge of both hyperfine structure constants
and magnetic moments. However, the anomalies can still be observed
from the measurements of the magnetic dipole constants in different
isotopes and the values of the $g$-factors
\cite{persson98,grossman99}. Deviations from Eq.
\ref{pointinteractionequality} are expressed in terms of the
hyperfine anomaly difference $_{87}\delta_{85}$:

\begin{equation}
\frac{A^{87}g^{85}_{I}}{A^{85}g_{I}^{87}}\cong1+_{87}\delta_{85},
\label{equation anomaly}
\end{equation}
with
$_{87}\delta_{85}=\epsilon^{87}_{BW}-\epsilon^{85}_{BW}+\epsilon^{87}_{BRCS}-\epsilon^{85}_{BRCS}$.
A $_{87}\delta_{85}\neq0$ indicates the presence of a hyperfine
anomaly.

\subsubsection{Breit-Crawford-Rosenthal-Schawlow effect}

\begin{figure}
\centering
  \includegraphics[width=2.7 in]{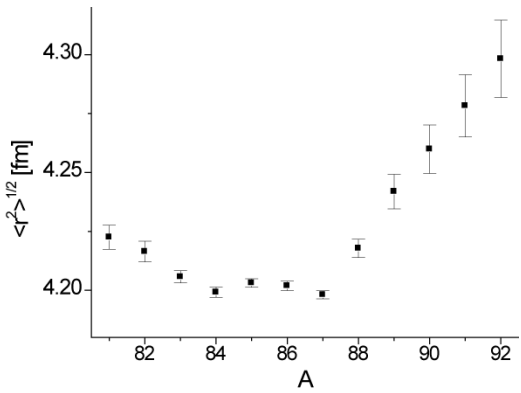}
  \caption{Plot of nuclear charge radius of rubidium as a function
  of atomic number. Adapted from Ref. \cite{angeli04}.}
  \label{charge distribution figure}
\end{figure}

The interaction between an electron and an atomic nucleus is
precisely described by the Coulomb potential when both of them are
far away from each other, no matter whether the nucleus is a point
or an extended source. For interactions that require the nucleus and
the electron to be at close distances, such as the hyperfine
interaction, an $1/r$ potential is no longer adequate. The correction to the electronic wave function due to the modified nuclear potential is known as the Breit-Crawford-Rosenthal-Schawlow correction.

Calculations of $\epsilon_{BRCS}$ take into consideration how the
charge is distributed over the nucleus. Rosenthal and Breit
considered for their calculation the charge to be on the surface of
the nucleus \cite{rosenthal32}. Schawlow and Crawford also calculated
the change of the wave function except they considered the charge
to be uniformly distributed in the nucleus \cite{crawford49}. Rosenberg and Stroke proposed later on a third model to improve the agreement between theory and experiment: a diffuse nuclear charge distribution \cite{rosenberg72}.  

The neutron and proton shells in rubidium determine the deformation as
well as the spatial distribution of the nuclear charge. The neutron
shell for $^{87}$Rb is closed at magic number $N=50$ making it
impervious to the addition and substraction of nuclear matter
\cite{thibault81,angeli04}. The substraction of two neutrons to form
$^{85}$Rb does not affect significantly the electric charge
distribution, and the electric potential, compared to the one from
$^{87}$Rb, remains the same (see Fig. \ref{charge distribution
figure}).

The expression of $\epsilon_{BCRS}$ for the uniformly charged sphere and charge on surface models is \cite{armstrong}:

\begin{equation}
\epsilon_{BCRS}=\frac{2(\kappa+\rho)\rho(2\rho+1)}{(2\kappa+1)(\Gamma(2\rho+1))^{2}}(\frac{pZr_{N}}{a_{0}})^{2\rho-1},
\label{equation BCRS correction}
\end{equation}
where $p$ is a constant of order unity, $\rho=\sqrt{\kappa^{2}-(Z\alpha)^{2}}$, $a_{0}$ and $\alpha$ are the Bohr radius and electromagnetic coupling constant, respectively,
r$_{N}$ is the nuclear radius, and $\kappa$ is related to the
electronic angular momentum through the equation $\kappa=1+J(J+1)-L(L+1)-S(S+1)$. Table \ref{Table BCRS correction} shows the value of the correction for a uniformly distributed  charge as well as the nuclear radius of each isotope employed in the calculation. 

\begin{table}[h]
  \leavevmode \centering
   \begin{tabular}{lccc}
                       &  r$_{N}$ [fm] & Ref.                    & $\epsilon_{BCRS}$ \\ \hline
   $^{85}$Rb &  4.2031(18)  & \cite{angeli04} & 0.0090835(34) \\
   $^{87}$Rb &  4.1981(17)  & \cite{angeli04} & 0.0090735(36) \\
   \end{tabular}
  \caption{Values of $\epsilon_{BCRS}$ and corresponding nuclear radius for both rubidium isotopes.}
  \label{Table BCRS correction}
\end{table}

Rosenfeld and Stroke propose a trapezoidal charge distribution to approximate their model. The interested reader should consult Ref. \cite{rosenberg72} for further explanation. All three models give relatively large   $\epsilon_{BCRS}$ ($\sim$1\%), however, the difference between
both isotopes for all models is very small: $\epsilon^{87}_{BCRS}-\epsilon^{85}_{BCRS}\sim10^{-5}$.

\subsubsection{Bohr-Weisskopf effect}

The interplay between nuclear magnetization with the magnetic field created by the atomic electrons causes the hyperfine splitting in atoms. A natural extension of hyperfine splitting measurements is to compare models of nuclear magnetism. 

Nuclear magnetization is described in terms of nuclear moments with the biggest contribution coming from the nuclear magnetic dipole moment. The assumption of a point magnetic dipole gives good agreement between calculations and experiment, however it does not provide the complete picture. Nuclear magnetization has a finite volume. The electron wavefunctions of levels with total angular momentum $J=1/2$ have a bigger overlap with the nucleus and are able to experience the subtle changes of the spatial distribution of the nuclear magnetization. This wave functions need to be modified to correctly account for the hyperfine splitting. 

The corrections $\epsilon_{BW}$ to the wave functions due to a finite magnetization distribution were first computed by Bohr and Weisskopf \cite{bohr50}. They assumed a uniformly distributed magnetization over the nucleus for their calculation with a predicted  $\epsilon^{87}_{BW}-\epsilon^{85}_{BW}$ that ranges between 0.11\% and 0.29\%. The BW correction roughly scales as \cite{bohr50}:

\begin{equation}
\epsilon_{BW}\sim(\frac{Zr_{N}}{a_{0}})(\frac{a_{0}}{2Zr_{N}})^{2(1-\sqrt{1-(Z\alpha)^2})}(\frac{r^2}{r_{N}^{2}})_{Av},
\label{equation BW correction}
\end{equation}
where the average is taken over the magnetization distribution, with $(r^{2}/r^{2}_{N})_{Av}=3/5$ for a uniform magnetization. For rubidium this gives a correction of the order of 0.2\%, however it is strongly dependent on spin and orbital states of the nucleons i.e. on the specifics of the nuclear magnetization. Stroke \textit{et. al.} performed the same calculation using a trapezoidal magnetization distribution \cite{stroke61}. Their results agree very well with experimental information extracted from the ground state; they calculate a hyperfine anomaly difference of 0.33\%. Both of this results are independent for the main quantum number of the valence electron \cite{kopfermann}.

The nuclear shell model predicts that the total magnetic dipole moment has contributions from both the proton and the neutron shell, each with orbital and spin angular momenta \cite{kopfermann}

\begin{equation}
\vec{\mu}=\sum_{i=n,p}(g^{eff}_{s,i}\vec{s}_{i}+g^{eff}_{l,i}\vec{l}_{i})\mu_{N},
\label{equation nuclear momenta}
\end{equation}
where $g^{eff}_{s}$ and $g^{eff}_{l}$ are the effective nuclear spin
and nuclear orbital gyromagnetic ratios, respectively, $\vec{s}$ and
$\vec{l}$ are the nuclear spin and nuclear orbital angular momenta
and the sum is taken over both shells. The $g$-factors have the
values $g^{eff}_{s}$=3.1(2) and $g^{eff}_{l}$=1.09(2)
\cite{yamazaki70}.

The magnetic dipole moment in rubidium comes almost entirely from the vector
addition of the orbital and spin angular momenta of a single
valance proton. The neutron shell is almost spherical for both
isotopes due to its closed shell structure and the contribution to
the angular momentum from the neutron shell is very small.

The lighter of the two isotopes, $^{85}$Rb, has the valence proton
in an almost degenerate $f$ orbital with its spin and orbital
momenta antialigned yielding a value of $I$=5/2. Adding two more
neutrons to the core shifts the energy level of the valence proton
to the nearby $p$ orbital and aligns both momenta giving the known
value of $I$=3/2. Table \ref{Table nuclear magnetic moments}
presents the theoretical prediction of the nuclear magnetic moment
using Eq. \ref{equation nuclear momenta} as well as the experimental
result. It is indeed remarkable that such a simple model reproduces
closely the experimental results, particularly for the closed nuclear shell structure of $^{87}$Rb.

\begin{table}[h]
  \leavevmode \centering
   \begin{tabular}{lcccccc}
                       &  Theory [$\mu_{N}$] & Experiment [$\mu_{N}$] & Ref. \\ \hline
   $^{85}$Rb & 2.00                            & 1.35298(10)                      &\cite{duong93} \\
   $^{87}$Rb & 2.64                            & 2.75131(12)                      &\cite{duong93} \\
   \end{tabular}
  \caption{ Theoretical and experimental values of the nuclear dipole moment for rubidium.}
  \label{Table nuclear magnetic moments}
\end{table}

Three main factors make the two stable isotopes of rubidium good candidates for observing the BW effect: First the different orientation of the nuclear spin of the valence proton with respect to the nuclear orbital angular momentum. Second, the small relative difference in nuclear charge deformation. Third, the change of orbital for the valence proton in the two isotopes.

\subsection{Two-photon spectroscopy}

\begin{figure}[b]
\leavevmode \centering
  \includegraphics[width=2.5in]{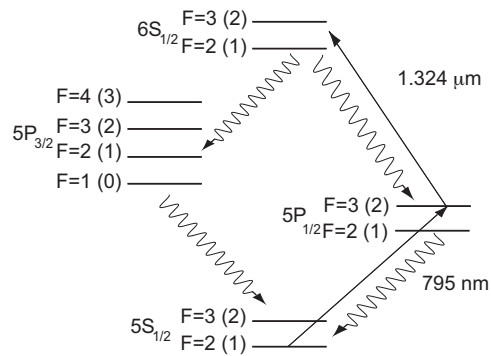}
  \caption{Energy levels relevant to our experiment (energy separations not drawn to scale). The numbers
  correspond to $^{85}$Rb ($^{87}$Rb). Straight arrows correspond to the excitation lasers, ondulated arrows to decays.}
  \label{complete energy levels}
\end{figure}

We use atomic laser spectroscopy to measure the hyperfine splitting
in two isotopes of rubidium. Parity requires a two photon electric dipole transition
to reach the 6S$_{1/2}$ state from the 5S$_{1/2}$ ground state. We
increase the probability of transition by using the 5P$_{1/2}$ level
as an intermediate step. We develop a theoretical model of the
two-photon transition that includes the main physical aspects of our atomic 
system (see Fig. \ref{complete energy levels}) based on a density
matrix formalism.

Our experimental setup consists of two counter propagating laser
beams going through a glass cell with rubidium vapor in a small
magnetic field. We lock the laser at 795~
nm on resonance, the middle step to the 5P$_{1/2}$ level, while we scan the 1.324 $\mu$m laser (from here on referred to as the 1.3~$\mu$m laser) over the 6S$_{1/2}$ level and
observe the absorption of the 795 nm laser. The system can be
modeled as a three level atom in which the on-resonance middle step
enhances the excitation to the final step and the counter
propagating laser beams help suppress the Doppler background (see
for example Ref. \cite{stenholm}). However, numerical simulations
show that we have to model our system as a five level atom to
include its main qualitative feature: optical pumping effects
increase the absorption of the 795~nm laser when the 1.3~$\mu$m
laser is on resonance.

\begin{figure}[t]
\leavevmode \centering
  \includegraphics[width=2.5in]{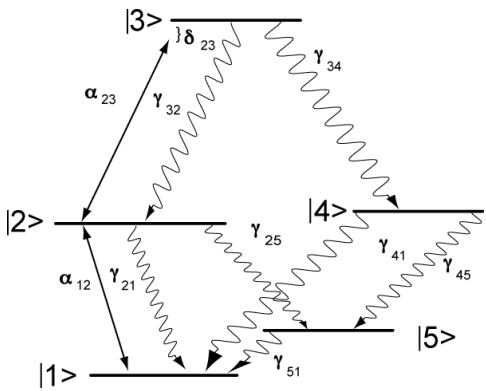}
  \caption{Energy level diagram of the theoretical model representing our system. The $\gamma_{i,j}$    corresponds to the decay rate between levels $|i\rangle$, $|j\rangle$, $\alpha_{i,j}$ is the Rabi frequency relating levels $|i\rangle$ and $|j\rangle$, and $\delta_{23}$ is the detuning from resonance of the exctitation laser between levels $|2\rangle$  and $|3\rangle$.}
    \label{figure bloch diagram}
\end{figure}

Figure \ref{figure bloch diagram} shows our simplified atomic model.
We have neglected the Doppler effects as well as the Zeeman
sublevels in order to keep the calculation as simple as possible
without losing the main qualitatively features of our system. Level
$|1\rangle$ represents the lower hyperfine state of the 5S$_{1/2}$
level while $|2\rangle$ is the upper hyperfine state of the
5P$_{1/2}$. The decay rate between the two levels is
$\gamma_{21}/2\pi=$ 6~MHz \cite{simsarian98}. We simplify the
hyperfine states of the 6S$_{1/2}$ level to just one level with
decay rate $\gamma_{32}/2\pi=$ 3.5~MHz \cite{gomez05b}. The ground
and intermediate levels are coupled by the Rabi frequency
$\alpha_{12}$ while the intermediate and the excited levels are
coupled by $\alpha_{23}$. The remaining two levels, $|4\rangle$ and
$|5\rangle$, represent all other decay channels out of the cascade
system and the upper hyperfine ground level, respectively. The
detuning between levels $|1\rangle$ and $|2\rangle$ is zero for our
experiment, but we let the detuning between levels $|2\rangle$ and $|3\rangle$ vary as $\delta_{23}$. The total population is normalized to one.

We are left with a set
of twenty five linear equations for the slowly varying elements of
the density matrix $\sigma_{nm}$ after using the rotating wave approximation.  These are
\begin{eqnarray*}
\lefteqn{\sum_{k}(\gamma_{kn}\sigma_{kk}-\gamma_{nk}\sigma_{nn})~+}&\\
  & & \frac{i}{2}\sum_{k}(\alpha_{nk}\sigma_{kn}-\sigma_{nk}\alpha_{kn})=0~for~n=m,\nonumber\\
\lefteqn{[i(\Omega_{nm}-\omega_{nm})-\Gamma_{nm})]\sigma_{nm}~+}\\
  & & \frac{i}{2}\sum_{k}(\alpha_{nk}\sigma_{km}-\sigma_{nk}\alpha_{km})=0~for~n\neq m,\nonumber
\end{eqnarray*}
where $\omega_{nm}=(E_{n}-E_{m})/\hbar$ is the transition frequency,
$\Omega_{nm}=-\Omega_{mn}$ is the laser frequency connecting the
levels. The damping rate is given by:
\begin{equation*}
\Gamma_{nm}=\frac{1}{2}\sum_k(\gamma_{nk}+\gamma_{mk}).
\end{equation*}

\begin{figure}[b]
\leavevmode \centering
   \includegraphics[width=3in]{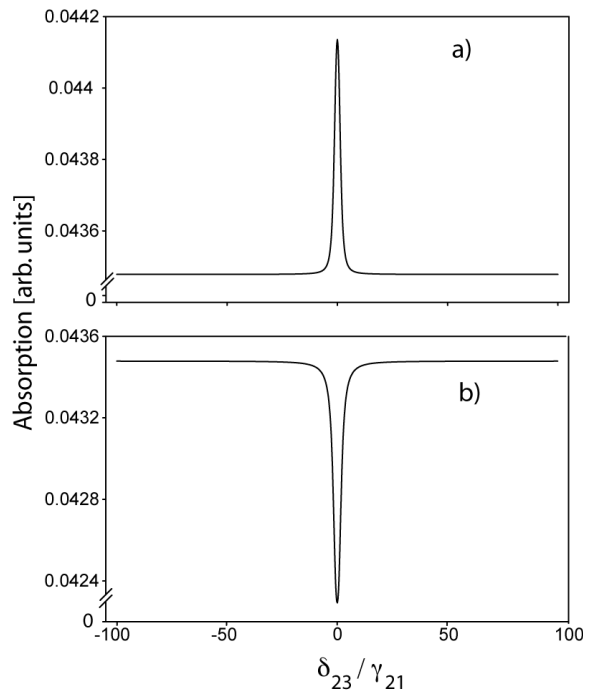}
  \caption{Numerical simulation of the absorption of the 795~nm laser as a function of
 the normalized detuning of the 1.3 $\mu$m laser to level $|3\rangle$ in units
of $\gamma_{21}$. Both plots have the same parameters except for the
ratio $\gamma_{41}/\gamma_{45}$. (a) Increase of absorption with
$\gamma_{41}/\gamma_{45}=2$. (b) Decrease of absorption with
$\gamma_{41}/\gamma_{45}=1/2$. }
  \label{absorption fig model}
\end{figure}

We solve for $\sigma_{12}$ leaving the detuning between levels
$|2\rangle$ and $|3\rangle$ ($\delta_{23}=\Omega_{23}-\omega_{23}$)
as a free parameter. We plot the negative of the imaginary part of
$\sigma_{12}$, which is proportional to the absorption of level
$|2\rangle$, as a function of $\delta_{23}$ for several different
sets of parameters. Our five level model reproduces the increase of
absorption observed as the second excitation goes into resonance.
This can be explained in the following way. The laser coupling
levels $|1\rangle$ and $|2\rangle$, in the absence of the second
excitation, pumps the atoms to level $|5\rangle$. On steady state
there will be little absorption due to a very small amount of atoms
being transferred from $|5\rangle$ to $|1\rangle$. By adding the
second excitation a new reservoir of ``fresh'' unexcited atoms
appears in level $|1\rangle$. Instead of falling to the
non-absorbing level $|5\rangle$, they travel to level $|3\rangle$
and then decay to the initial ground state level through level $|4\rangle$. These
``fresh'' atoms will add to the ground state population and increase
the absorption.

Figure \ref{absorption fig model} shows samples of our simulation. We
have plotted the absorption of the laser connecting levels
$|1\rangle$ and $|2\rangle$ as a function of the detuning of the
second laser. Figure \ref{absorption fig model} (a) shows how the
absorption increases as the second laser goes on resonance while
Fig. \ref{absorption fig model} (b) shows a decrease. Both plots
have the same model parameters except for the ratio
$\gamma_{41}/\gamma_{45}$. This ratio determines whether the atom
will be lost or return to the cycle. A ratio bigger than one pumps
atoms preferentially to level $|1\rangle$ rather than level
$|5\rangle$ which constitutes a fresh reservoir of excitable atoms.

\section{Measurement of the hyperfine splitting}

\subsection{Apparatus}

\begin{figure}[b]
\leavevmode \centering
    \includegraphics[width=2.5in]{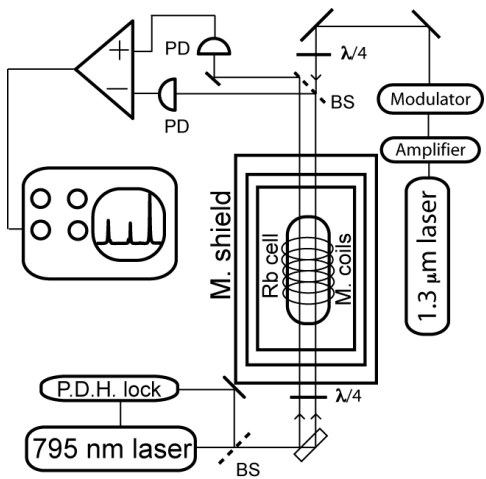}
    \caption{Block diagram of the experiment. Key for figure PD: photodiode, P.D.H.: Pound-Drever-Hall, M: magnetic, BS: beamsplitter.}
    \label{block diagram figure}
\end{figure}

We use a Coherent 899-01 Titanium Sapphire (Ti:sapph) laser with a
linewidth of better than 500 kHz tuned to the D1 line at 795 nm for
the first step of the transition. A Pound-Drever-Hall (PDH) lock to
the F=1(2)$\rightarrow$F=2(3) transition in $^{87}$Rb $(^{85}$Rb) in
a separate glass cell at room temperature stabilizes the linewidth
and keeps the 795~nm laser on resonance. An HP 8640B signal
generator acts as the local oscillator for the lock. The 795~nm laser remains on resonance for about 40 minutes, much longer than the time it takes to record a single experimental trace. 

A grating narrowed diode laser at 1.3 $\mu$m with a linewidth better than 500
kHz excites the second transition. We scan the frequency of the 1.3
$\mu$m laser with a triangular shaped voltage ramp from a
synthesized function generator at 4 Hz applied to the piezo control
of the grating and monitor its frequency with a wavemeter
with a precision of $\pm$0.001 cm$^{-1}$. A fiber-coupled
semiconductor amplifier increases the power of the 1.3 $\mu$m laser
before it goes to a large bandwidth ($\approx$10 GHz) Electro-Optic
Modulator (EOM). Another HP 8640B modulates this EOM. Fig.
\ref{block diagram figure} shows a block diagram of the experimental
setup.

A thick glass plate splits the 795 nm laser beam into two
copropagating beams before going to the glass cell. The power of each beam is approximately
10~$\mu$W with a diameter of 1 mm. We operate in the low intensity
regime to avoid power broadening, differential AC stark shifts and line splitting effects such as
the Autler-Townes splitting. Both beams are circularly polarized by a $\lambda/4$ waveplate. A
counter propagating 1.3 $\mu$m laser beam with a power of 4 mW and
approximately equal diameter overlaps one of the 795~nm beams. The
lasers overlap to a precision of better than 1 mm along 75 cm. The
cell resides in the center of a 500-turn solenoid that provides a
magnetic field of 7.4~Gauss/A contained inside a three layered
magnetic shield to minimize magnetic field fluctuations
\cite{burt02}. The middle layer has a higher magnetic permeability
to avoid saturation effects. The solenoid is 70 cm long and has a
diameter of 11.5 cm. We operate under a weak magnetic field
($B\approx$1 Gauss) to work in the Zeeman linear regime.

After the glass cell an independent photodiode detects each 795~nm
beam. The outputs of the detectors go to a  differential amplifier
to reduce common noise. A digital oscilloscope records the output
signal for different values of modulation, polarization and magnetic
field and averages for about three minutes. The order in which the
absorption profiles are recorded is random. During the experimental
runs we monitor the current going to the solenoid that provides the
quantization axis. A thermocouple measures the changes in
temperature inside the magnetic shield (24 \textcelsius) to within
one degree. The optical attenuation for the D1 line at line center is 0.4 for
$^{85}$Rb and about three times less for $^{87}$Rb.

\subsection{Method}

\begin{figure}[t]
\leavevmode \centering
  \includegraphics[width=3in]{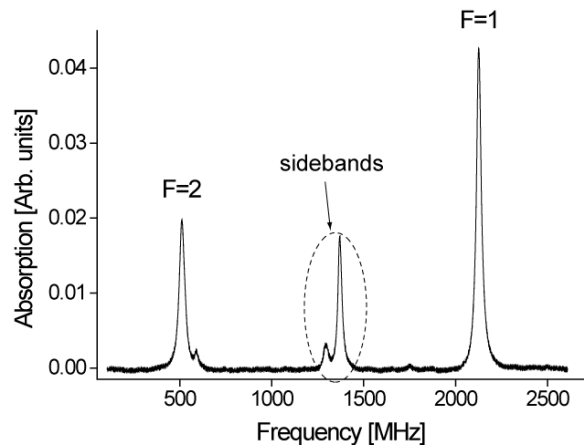}
  \caption{Absorption profile of the 6S$_{1/2}$, F=1 and F=2 hyperfine states of $^{87}$Rb with sidebands. The big sideband belongs to the F=1 peak. The small feature on the side of the F=2 peak corresponds to the second sideband of the F=1 peak. The glass cell is in a magnetic field of 0.37 G.}
  \label{figure whole scan}
\end{figure}

\begin{figure}[b]
\leavevmode \centering
  \includegraphics[width=3in]{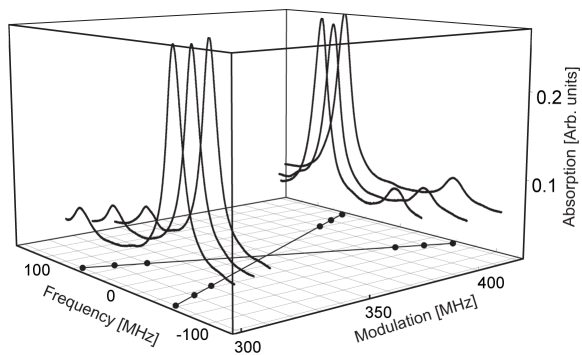}
  \caption{Experimental traces that illustrate sideband crossing for
  $^{85}$Rb. The larger resonance corresponds to the $F=2$ level, the smaller one to the $F=3$ level of the 6S$_{1/2}$ state. The dots correspond to the center of the profiles, the point where both lines cross corresponds to half the hyperfine separation. }
  \label{figure cross3d}
\end{figure}

We modulate the 1.3 $\mu$m laser to add sidebands at an appropriate frequency with a modulation
depth (ratio of sideband amplitude to carrier amplitude) that ranges between 1 and 0.1. The sidebands appear in the absorption profile at a distance equal to the modulation from the
main features and work as an \emph{in situ} scale (see Fig. \ref{figure whole scan}). We measure their separation as a function of the modulation for values bigger and smaller than half the hyperfine
splitting. We interpolate to zero separation to obtain half the hyperfine splitting (see Fig. \ref{figure cross3d}). This technique transfers an optical frequency measurement to a much easier
frequency measurement in the RF range.

The size of the main peaks depends on the coupling strength between
transitions; the size of the sidebands (as compared to the
main peaks) will be determined by the strength of the
transition and also on the number of sidebands simultaneously on or close
to resonance. We observe under normal experimental conditions that the laser sidebands
are both close to resonance (the lower frequency sideband to the 6S$_{1/2}$
$F$=1 and the upper one to the $F$=2 transition) when the carrier is
around the half point of the splitting. The stronger of the
transitions ($F$=1) depopulates the 5P$_{1/2}$ $F$=2 level leaving
only a few atoms to excite with the upper sideband, hence the
smaller transmission peak for the sideband corresponding to $F$=2.

\begin{figure}[t]
\leavevmode \centering
    \includegraphics[width=3in]{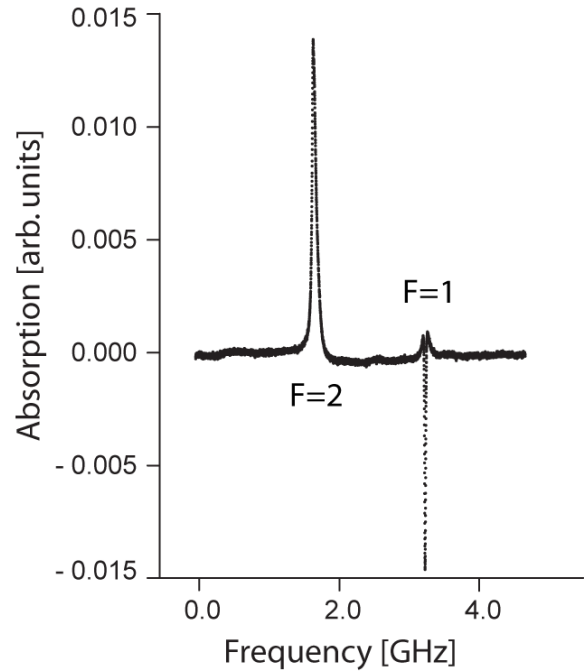}
    \caption{Experimental trace of absorption of the 795~nm laser for $^{87}$Rb showing both increase and decrease of absorption due to optical pumping.}
    \label{figure optical pumping}
\end{figure}

We have also observed a much richer atomic behavior by changing the
laser intensities, polarizations and magnetic field environment of
the glass cell. Optical pumping moves the atomic population from one
level to another quite efficiently. This is manifest in how the
peaks change in magnitude or just switch from an increase of
absorption to a decrease (see Fig. \ref{figure optical pumping})
just as our very simple theoretical model predicts. These effects
point out that a careful control of the environment is necessary for
a successful realization of the experiment.

The transfer of population by specific selection of polarization and
magnetic environment can also be used to obtain a better
experimental signal. There are several options to reach the
6S$_{1/2}$ level. From the ground hyperfine states we can do $\Delta
F=0,\pm 1$ transitions. We find that doing the two step excitation
in either a $\sigma^{+}:\sigma^{-}$ or $\sigma^{-}:\sigma^{+}$
polarization sequence for the 795 nm and 1.3~$\mu$m lasers,
respectively, with a $\Delta F=1$ for the first step increases the
amplitude of the signal. By choosing this polarization sequence we
increase the probability of the atom going to the excited state and
avoid placing it in a non-absorbing state \cite{budker04}.

We place the rubidium cell in a uniform magnetic field collinear
with the propagation vectors of both lasers. The magnetic field
provides a quantization axis as well as a tool to probe systematic
effects. The hyperfine separation is now dependent on the magnetic
field strength and the alignment with the laser. We measure the
hyperfine splitting for different values of the magnetic field and
polarization making sure that the above polarization sequence is
always satisfied. We extract the value of the splitting at zero
magnetic field from a plot of hyperfine splitting as a function of
magnetic field.

\subsection{Systematic effects and results}

\begin{figure}[b]
\leavevmode \centering
   \includegraphics[width=3in]{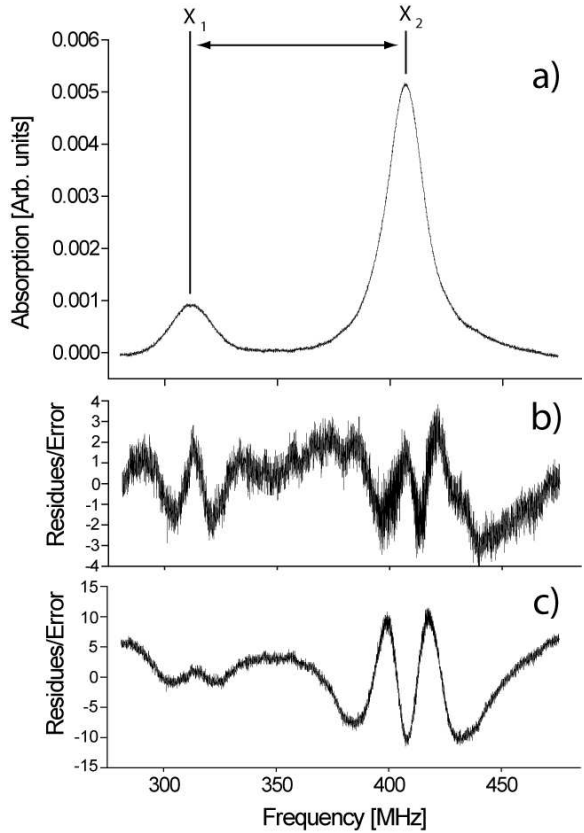}
  \caption{(a) Scan of the sidebands of the 6S$_{1/2}$, $F$=1 and $F$=2 hyperfine states of
  $^{87}$Rb. The fits are not shown for
  clarity. (b) Normalized residuals of the Lorentzian fit, the reduced $\chi^{2}$ is 2.13. (c) Normalized residuals of the Gaussian fit, the reduced $\chi^{2}$ is 23.13.}
  \label{figure sidebands and fits}
\end{figure}

We study the contributions of several systematic effects that can
influence the hyperfine separation measurement. We analyze the peak
shape model for the non-linear fit to obtain the separation of the
centers of the profiles, scan width and scan rate of the 1.3
$\mu$m laser, power of the 795~nm and 1.3 $\mu$m lasers, optical
pumping effects, magnetic field effects, and temperature.

A)\emph{Peak shape model and non-linear fit.} The absorption of a Doppler-broadened two level system as a function of laser detuning is a Voigt profile. When a multilevel system is considered it is not trivial to write down the functional form of the absorption of any of the lasers interacting with the system (see for example Refs. \cite{marquardt96,banacloche95}). We fit the experimental data to Voigt, Lorentzian and Gaussian functions to find the line centers and compare the results for consistency.

We use the non-linear fit package of ORIGIN\texttrademark~ to fit the above mentioned profiles to search for model-dependent systematics. ORIGIN\texttrademark~ uses a Levenberg-Marquardt algorithm to minimize the residuals given a specified error. The program has been
used in the past by our group to obtain high precision lifetime measurements \cite{gomez05a,gomez05b}. We use the resolution limit of the 8 bit analog to digital converter of the scope for these
calculations which corresponds to 0.5\% of the total scale used. Lorentzian and Gaussian fits have three variable parameters to fit for each peak which correspond to the FWHM, the line center, the
area under the curve plus a single offset for both peaks. Voigt profiles have an extra parameter which corresponds to the temperature of the sample. ORIGIN\texttrademark~ gives the error of
each parameter which depends on the quality of the data. 

Voigt profiles are in very good agreement with the lineshape. The fit yields the low temperature limit of the Voigt profile i.e. a Lorentzian, and hence is in agreement with the linecenter extracted using a Lorentzian profile. This is expected since the contribution of the Doppler effect on the resonance lineshape should be minimized by the counter propagating laser setup and by an expected group velocity selection arising from the the two-step excitation process (``two-color hole burning''). The 795 nm laser will only interact with a small number of group velocities; these groups will be the only ones that will be excited to the 6S$_{1/2}$ level by the 1.3 $\mu$m laser. Line centers extracted from Gaussian fits agree with results from the above mention profiles but decay too fast for frequencies far away from the centers.  We also fit the data to a convolution of Lorentzian profiles with a rectangular transmission function and an exponential of a Lorentzian to search for systematic errors and to understand better our residues.

All peak shape models give line centers consistent among themselves. All of them have similar structures in the residues within the line width of the resonances (see Fig. \ref{figure sidebands and fits}). We have determined that these features come about from the high sensitivity from deviations from a perfect fit that a difference of two peak profiles has. In other words, by taking the residues we are effectively taking the derivative of a peak profile that will be as sensitive as sharp the linewidth is. To further verify this we take the numerical derivative of the data to search for residual structure that might change our measurement. We fit a straight line to the data that lies within the linewidth and extract when the line crosses zero. The results are consistent with the fits. Close analysis of the derivative in this region reveals no structure.

Of the fitted functions Lorentzians yield the smallest $\chi^{2}$. The fitting error of the line centers for all  our data for Lorentzian fits range  between 15 kHz and 30 kHz. We quote the average of all the fitting errors of our data in Table \ref{table uncertainty}. Fig. \ref{figure sidebands and fits} shows a zoom of the sidebands as well as the residues for a Lorentzian and Gaussian fits. We extract the line centers with both models; the difference in separation for both models is in this case $|x_{1}-x_{2}|_{Lorentzian}-|x_{1}-x_{2}|_{Gaussian}=0.35(68)$~MHz. The reduced $\chi^{2}$ of the non-linear Lorentzian fit for all our data ranges between 1 and 10 depending on the noise of the signal with a $\chi^{2}$ average of 2.4 over twenty fits. We do not observe changes in the splitting that depend on the frequency range fitted
around the resonances. 

The relative angle between both copropagating lasers induces a systematic shift on the absolute frequency the atoms observe due to the appearance of the $\vec{v}\cdot\vec{k_{i}}$ dependence on absorption where $\vec{v}$ is the velocity of the atom and $\vec{k_{i}}$ is the wave vector for either laser. This angle dependence on the Doppler shift for our system is almost the same for both our lasers  since the cosine of the angle between them differs from one by one part in $10^{5}$. Furthermore, any residual effect is minimized since we measure frequency differences.
 
Just like the line shape, analytic expressions for the linewidth are difficult to write down. We perform a numerical simulation of our five level system presented in subsection \emph{D} of the theoretical background in the presence of a room temperature velocity distribution. The resonances show linewidths of the order of 30-40 MHz which are in very good agreement with experimental results.

Distortions of the lineshape i.e. asymmetries, depend on the detuning of  the 795~nm laser from resonance. These can induce unwanted systematic errors to the measurement. Numerical simulations show, following Ref \cite{grossman00a}, that the separation of the hyperfine splitting depends negligibly on the detuning from the D1 line. Nevertheless, we look for any asymmetries in the peaks themselves and dependence on the direction of scan during experimental runs. No correlation with these effects is found.

We interpolate to zero from a plot of distance between the center of
the sidebands \textit{vs.} the modulation frequency to obtain half
the hyperfine separation. The linear regression coefficients in this
plots differ from one at the most in 2 parts in $10^{4}$. Typical
errors for the crossovers amount to about 200 kHz.

\begin{figure}[t]
    \includegraphics[width=3in]{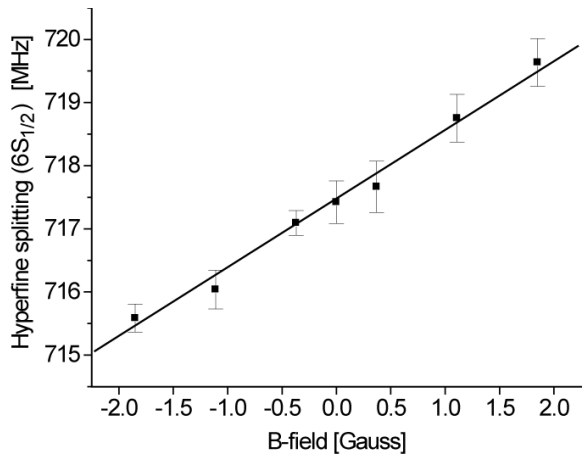}
    \caption{Zeeman plot of the hyperfine separation of the 6S$_{1/2}$ level of $^{85}$Rb  with both lasers circularly polarized to better than 95\% and linear fit.}
    \label{zeeman no jump}
\end{figure}

B)\emph{Scan and linearity of 1.3 $\mu$m laser.} Non-linearities in
the piezo driving the feedback grating, hysteresis effects as well
as a slow thermal drift on the 1.3 $\mu$m laser can generate
undesired systematics in the measurement. We look for
non-linearities by sending the voltage monitor of the piezo to a
digital scope with an 8-bit resolution during the experimental runs
as well as monitor the absorption peaks for asymmetries. Comparison
between absorption peaks for both types of scan (low to high
frequency and vice versa) reveals no systematic effects. Analysis of
the long term drift of the 1.3 $\mu$m laser shows a stability of
better than 100 kHz over a 5 min. period which is longer than the
time we need to take a single experimental absorption signal.

C)\emph{Power of the 795 nm and 1.3 $\mu$m laser.} We look for
systematic dependence on the hyperfine splitting on the power of both lasers. We change the
power of the 795 nm laser from 4 $\mu$W to 40 $\mu$W while keeping the power of the 1.3 $\mu$m laser constant. Low signal to noise ratio and the observation of the Autler Townes splitting
determine the lower and upper boundaries of this interval,
respectively. 

The Autler Townes effect predicts a splitting of the middle energy level by the on-resonance first step in a three level system that is proportional to the square root of its intensity \cite{delsart78}. For our typical experimental conditions the splitting should be less than 4 MHz, very small to be resolved with the observed linewidths of the atomic resonances ($\sim40$ MHz).

The 1.3~$\mu$m laser operates very close to its
maximum power on normal experimental conditions. The power is
distributed among the sidebands and the main carrier depending on
the modulation depth. We gradually decrease the power of the 1.3
$\mu$m to half its operating value to detect any dependence on the
power. We observe no correlation.

D)\emph{Optical pumping effects and magnetic field.} Optical pumping
effects are the most delicate of all the systematic effects. Both
laser beams are carefully polarized using appropriate $\lambda/4$
waveplates and their polarization checked with a rotating polarizer
in front of a detector to better than 95\%. The polarization of the lasers as well as
their alignment with the magnetic field determine the relative size
of the peaks ($m_{F}$ sublevels) that form the resonances of the
6S$_{1/2}$ hyperfine levels. Comparison of absorption profiles for a
set polarization sequence for different values of the magnetic field
gives qualitative information of the alignment between the magnetic
field and the lasers. The positive and negative magnetic field orientations in a perfectly symmetric situation, after a switch of polarization sequence, should yield the same absorption profile. For
everyday experimental conditions (around 1 G) we observe no
difference between positive and negative magnetic field directions.
We see broadening of the profiles at magnetic fields twenty times larger
but no asymmetries. Differences start appearing at around 85 Gauss
which suggests good alignment between the lasers and the magnetic
field as well as good control on polarization of both lasers.

The hyperfine separation \textit{vs.} magnetic field plot provides
more quantitative information. Eq. (3) states that the plot should
be linear with no discontinuities as we change the value of the
magnetic field from positive to negative. Our plots show a smooth
transition between negative and positive values of the magnetic
field within experimental error. Fig. \ref{zeeman no jump} shows a
sample of our data when both lasers are circularly polarized to
better than 95\%.

We monitor the current of the coil generating the magnetic field to
detect any fluctuation in the intensity of the field. We observe
small current noise that manifest into fluctuations at most of the order of
milligauss.

E)\emph{Temperature.} We analyze the position of the absorption
peaks as a function of temperature of the cell to check for related systematic
effects such as collision shifts for both isotopes. The
temperature of the glass cell is increased from room temperature (23
\textcelsius) up to 40 \textcelsius\ using a heat tape wrapped
around it. While recording data we turn off the heating tape to
avoid stray magnetic fields generated by the current going through
it. The temperature of the glass cell is monitored with a
thermocouple inside the magnetic shield with an accuracy of one
degree celcius. No dependency on temperature is found.

We have concluded after close analysis of these studies that, to the
accuracy of our measurement, Gaussianly distributed statistical
fluctuations dominate our experiment (see Table \ref{table
uncertainty}). The statistical error in the hyperfine splitting, as
stated by the standard error of the mean, is 110 kHz for $^{85}$Rb
and 167 kHz for $^{87}$Rb. 

\begin{figure}[b]
\leavevmode \centering
   \includegraphics[width=3in]{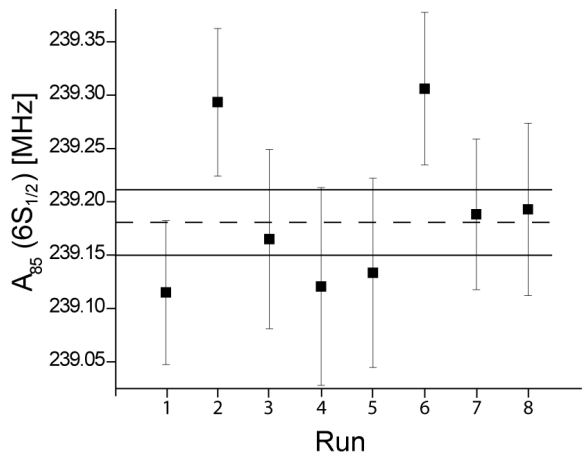}
  \caption{Results of different runs of the magnetic dipole constants of the 6S$_{1/2}$ state of $^{85}$Rb.
  The dashed line corresponds to the mean, the solid lines to the 1-$\sigma$ error.}
  \label{A85 dispersion}
\end{figure}

Figure \ref{A85 dispersion} shows the
values of the magnetic dipole constant for $^{85}$Rb for all
experimental runs of this work. The final result of each run is
determined by an interpolation to zero magnetic field as a function
of the current in the solenoid.

\begin{table}[t]
  \leavevmode \centering
   \begin{tabular}{lcc}
   Systematic effects       &$\nu^{85}_{HF}$[MHz]&$\nu^{87}_{HF}$[MHz]\\ \hline
   Optical pumping effects &        $\leq$ 0.016   & $\leq$ 0.029 \\
   Power of 795 nm laser   &        $\leq$ 0.020   & $\leq$ 0.005  \\
   Power of 1.3 $\mu$m laser  &     $\leq$ 0.011   & $\leq$ 0.011  \\
   Atomic density          &        $\leq$ 0.020   & $\leq$ 0.010  \\
   Non linear fit          &        $\leq$ 0.028   & $\leq$ 0.023  \\
   B-field fluctuations    &        $\leq$ 0.015   & $\leq$ 0.025  \\\hline
   Total Systematic        &        $\leq$ 0.047   & $\leq$ 0.047  \\\hline
   Statistical error       &        0.100          & 0.160  \\
   \textbf{TOTAL}           &        \textbf{0.110 }         & \textbf{0.167}  \\

   \end{tabular}
  \caption{Error budget for the hyperfine splitting measurement}
  \label{table uncertainty}
\end{table}

Table \ref{table final results} contains the measurements of the
hyperfine splitting of the 6S$_{1/2}$ level as well as the
corresponding values of the magnetic dipole constants for both
isotopes.

The precision of our data allows us to observe a hyperfine anomaly.
We use the values of Ref. \cite{stone05} for the ratio
$g^{85}_{I}/g^{87}_{I}=0.295055(25)$. This is consistent with the
experimental values of Ref. \cite{duong93}. Using this value and our
experimental results in Eq. \ref{equation anomaly} we obtain a value
for the hyperfine anomaly difference of
$_{87}\delta_{85}$=-0.0036(2). This is less than one percent
difference, well beyond the current MBPT theoretical calculation
accuracy of the hyperfine splittings.

\begin{table}[h]
  \leavevmode \centering
   \begin{tabular}{lcc}
                 & $^{85}$Rb [MHz] & $^{87}$Rb [MHz]\\\hline
   $\nu_{HF}$    &   717.54(10)    & 1615.32(16)     \\
   $A$           &   239.18(03)    & 807.66(08)      \\\hline
   \end{tabular}
  \caption{Hyperfine splittings ($\nu_{HF}$) and magnetic dipole constants for the 6S$_{1/2}$ level.}
  \label{table final results}
\end{table}

\section{Comparison with theory}

\begin{figure}
\leavevmode \centering
   \includegraphics[width=3in]{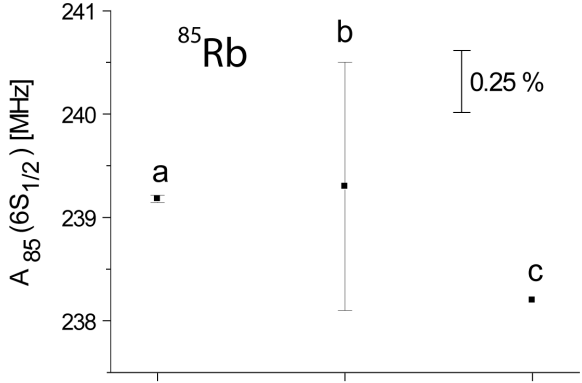}
  \caption{Comparison between experimental and theoretical results of the magnetic dipole constant of the 6S$_{1/2}$ state of $^{85}$Rb. The points labeled ``a'' and ``b'' correspond to our
  work and the work of Ref. \cite{gupta73}, respectively. Point ``c'' corresponds to the
   theoretical prediction of \cite{safronova99}.}
  \label{theory and experiment 85}
\end{figure}

We compare in Figs. \ref{theory and experiment 85} and \ref{theory
and experiment 87} the results from this experiment with the
previous experimental results of Gupta \textit{et al.}
\cite{gupta73} and the theoretical predictions of Safronova
\textit{et al.} \cite{safronova99}. The hyperfine anomalies are
still not within reach of MBPT so the value of $^{85}$Rb comes from
considering no hyperfine anomaly.

\begin{table}[h]
  \leavevmode \centering
   \begin{tabular}{l|ccc}
                 &  SD [MHz] & SDpT [MHz] & Experiment [MHz]\\\hline
   5S$_{1/2}$    &  642.6    & 1011.1     & 1011.910813(2) \cite{arimondo77}\\
   5P$_{1/2}$    &  69.8     & 120.4      & 120.499 (10) \cite{barwood91} \\
   6S$_{1/2}$    &  171.6    & 238.2      & 239.18(3) (this work)\\
   6P$_{1/2}$    &  24.55    & 39.02      & 39.11(3) \cite{marian04,marian05}\\
   7S$_{1/2}$    &  70.3     & 94.3       & 94.658(19) \cite{chui05} \\

   \end{tabular}
  \caption{Single Double (SD) and partial triple (SDpT) excitation calculated from \textit{ab intio} MBPT in Ref.  \cite{safronova99} and experiment magnetic dipole constants for the first $J$=1/2
   levels in $^{85}$Rb.}
  \label{table theory and experiment}
\end{table}

Table \ref{table theory and experiment} shows the values of the
magnetic dipole constants using relativistic MBPT \cite{safronova99}
with single double (SD) and single double partial triple (SDpT) wave
functions and values extracted from measurements of the hyperfine
splitting in other electronic states currently in the literature for
$J$=1/2 \cite{arimondo77,barwood91,marian04,marian05,chui05}. We
have not been able to find in the literature values for higher
levels with adequate precision to include them in the figure. The
agreement of the theory with the experiment, for $J$=1/2 levels, is
well within the 1\% level. The SDpT relativistic wave functions do
indeed improve the accuracy of the calculations of the single double
wave functions.

\section{Conclusions}

We have measured the hyperfine splittings of the 6S$_{1/2}$ level of
$^{85}$Rb and $^{87}$Rb to a precision of 103 ppm and 153 ppm,
respectively. Our measurement is consistent with and decreases the
uncertainty of the past measurements  \cite{gupta73} by a factor of 63 for $^{87}$Rb
and by a factor of 30 for $^{85}$Rb.

\begin{figure}[b]
\leavevmode \centering
   \includegraphics[width=3in]{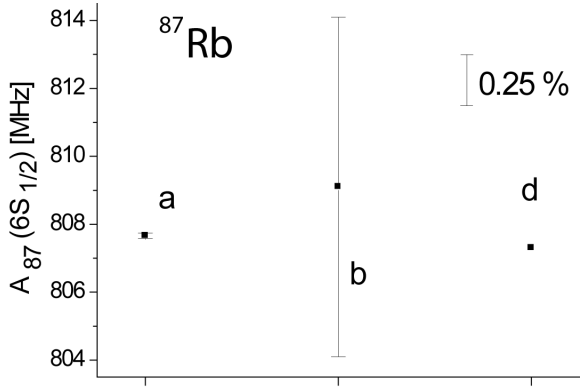}
  \caption{Comparison between experimental and theoretical results of the magnetic dipole constant of the 6S$_{1/2}$ state of $^{87}$Rb. The points labeled ``a'' and ``b'' correspond to our
  work and the work of \cite{gupta73}, respectively. Point ``d'' corresponds to the
   theoretical prediction of Ref. \cite{safronova99}. The value for $^{ 87}$Rb is obtained from the theoretical prediction for $^{85}$Rb by
   considering the hyperfine anomaly as nonexistent.}
  \label{theory and experiment 87}
\end{figure}

\begin{figure}
\leavevmode \centering
   \includegraphics[width=3in]{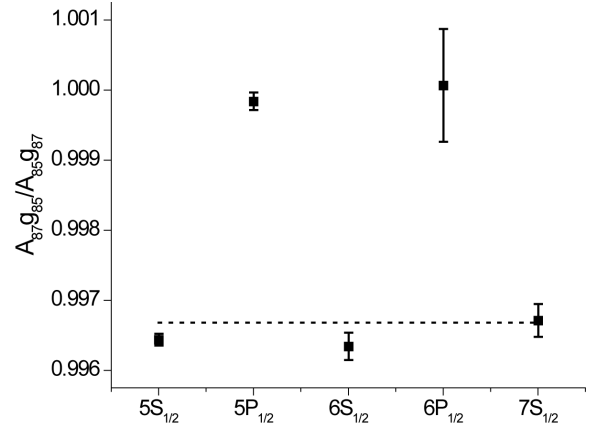}
  \caption{Hyperfine anomalies of other atomic levels of rubidium along with the value calculated
  in this measurement. The dashed line corresponds to the theoretical prediction for a diffuse magnetization distribution. See text for references.}
  \label{figure hyperfine anomaly}
\end{figure}

We are able to extract the hyperfine anomaly with our experimental
data and show that precision measurements
of the hyperfine structure in atomic states with different radial
distributions can give information on the nuclear magnetization
distribution. The hyperfine anomaly difference we extract for the 6S$_{1/2}$ is $_{87}\delta_{85}=-0.0036(2)$. The difference in the anomalies is indeed a factor of thirty larger than the expected BCRS
contribution and it comes from the BW effect. Fig. \ref{figure hyperfine anomaly} shows that the
anomaly measured with the nS$_{1/2}$ levels is the same independent
of the principal quantum number as well as the smaller deviation
from the point interaction, if any, for the nP$_{1/2}$ levels
\cite{perez07,arimondo77,barwood91,marian04,marian05,chui05}. Table \ref{table hyperfine anomaly differences}
shows the hyperfine anomaly differences for the first $J=1/2$
levels.

The plot and the table confirm that it is the electronic angular
momentum and not the principal quantum number that determines the
Bohr-Weisskopf effect and the bigger anomaly of the nS$_{1/2}$
levels due to their larger electronic density near the nucleus
\cite{bohr50}. These new measurements invite new calculations of
atomic properties and constrain nuclear calculations. As the nuclear
charge and magnetization distribution are better understood they
will further test and refine the calculations which are of crucial
importance for parity non-conservation experiments.

\begin{table}
  \leavevmode \centering
   \begin{tabular}{lcc}
                        & $_{87}\delta_{85}$ \\\hline
   5S$_{1/2}$ & -0.00356(8)  \cite{arimondo77}\\
   5P$_{1/2}$ & -0.0001(1) \cite{barwood91} \\
   6S$_{1/2}$ & -0.0036(2) (this work)\\
   6P$_{1/2}$ & 0.0000(8) \cite{marian04,marian05}\\
   7S$_{1/2}$ & -0.0032(2) \cite{chui05} \\\hline
      \end{tabular}
  \caption{Hyperfine anomaly differences $_{87}\delta_{85}$ for the first $J$=1/2 levels
  in rubidium.}
  \label{table hyperfine anomaly differences}
\end{table}

\acknowledgments Work supported by NSF. A.P.G. would like to thank
G. D. Sprouse for insightful discussions on nuclear physics and R.T. Willis for assistance with numerical simulations of the  atomic system.

\end{document}